\documentclass{ims9x6}
\allowdisplaybreaks
\usepackage{cite}

\DeclareMathOperator{\Tr}{Tr}
\newcommand{\norm}[1]{\left\lVert #1 \right\rVert}
\def\sh{\mathrm{sh}}
\def\ch{\mathrm{ch}}
\def\wt{\widetilde}
\def\cE {\mathcal{E}}
\def\cF {\mathcal{F}}
\def\cG {\mathcal{G}}
\def\cH{\mathcal{H}}
\def\cL{\mathcal{L}}
\def\cU{\mathcal{U}}
\def\cN{\mathcal{N}}
\def\C {\mathbb{C}}
\def\H {\gH}
\def\R {\mathbb{R}}
\def\T {\mathbb{T}}
\def\ph{\varphi}
\def\d{{\rm d}}  
\def\i{{\rm i}}  
\newcommand{\ao}{\mathfrak{a}}
\newcommand{\gH}{\mathfrak{H}}
\newcommand{\bH}{\mathbb{H}}
\newcommand{\product}[2]{\ensuremath{\left\langle #1, #2 \right \rangle}}
\newcommand{\dGamma}{{\ensuremath{\rm d}\Gamma}}

\makeatletter
\renewcommand{\ps@plain}{%
  \renewcommand{\@oddhead}{\hfil\footnotesize%
    \raisebox{30pt}[0pt][0pt]{\parbox{300pt}{\centering%
      A contribution to the Proceedings of the\\{}%
      Workshop on Density Functionals for Many-Particle Systems\\{}
      2--29 September 2019, Singapore}}\hfil}%
  \renewcommand{\@evenhead}{\@oddhead}%
  \renewcommand{\@oddfoot}{\hfil\footnotesize%
        \raisebox{-8pt}[0pt][0pt]{\thepage}\hfil}%
  \renewcommand{\@evenfoot}{\@oddfoot}%
}
\makeatother
\renewcommand{\trimmarks}{\relax}

\begin{document}

\chapter{\uppercase{Dynamics of interacting bosons:\\%
                                   A compact review}}

\markboth{M. Napi\'orkowski}%
{Dynamics of interacting bosons: A compact review}

\author{Marcin Napi\'orkowski}
\address{Department of Mathematical Methods in Physics\\ %
  Faculty of Physics, University of Warsaw\\ %
  Pasteura 5, 02-093 Warszawa, Poland\\ 
 marcin.napiorkowski@fuw.edu.pl}

\begin{abstract}
  The success of the Gross--Pitaevskii and Bogoliubov theories in the
  description of large systems of interacting bosons led to a substantial
  effort into rigorously deriving these effective theories.
  In this work we review the related literature in the context of
  dynamics of large bosonic systems.  
\end{abstract}

\section{Introduction}
\subsection{Setup}
Bose--Einstein Condensation (BEC) is a phenomenon that occurs in systems of
bosonic particles where, at sufficiently low temperatures, a macroscopic
fraction of the particles  starts to occupy a single quantum state.
The idea of BEC dates back to the works of Bose and Einstein in 1924
\cite{Bose-24,Einstein-24} in which they analysed non-interacting systems.
Experimentally BEC has been produced only in 1995 by Cornell and Wieman, and
Ketterle \cite{CorWie-95,Ketterle-95}.
Since then, fundamental questions in the rigorous understanding of
condensation and fluctuations around the condensate in interacting systems
have remained unanswered.
Some of them are essential for the understanding of interesting quantum
effects that can be observed even on macroscopic scales (such as, for example,
superfluidity).  

In this work we will review the existing results concerning the dynamics of
Bose--Einstein condensates.
The physical picture we have in mind is as follows.
Let $\Psi_{N,0}$ be the ground state of $H_N^V$ given by 
\begin{equation} \label{eq:HN-V} 
  H_N^V= \sum\limits_{j = 1}^N \bigl(-\Delta_{x_j}+V(x_j)\bigr)
  + \frac{1}{N-1} \sum\limits_{1 \le j < k \le N} {w_N(x_j-x_k)}.
\end{equation}
Here $V\in L^\infty_{\rm loc}(\R^3,\R)$, satisfying $V(x)\to \infty$ as $|x|\to
\infty$, is an external potential (which models the trapping of the particles
while they are cooled during the experiment) and $w_N$ is the inter-particle
potential that describes their interactions (and could a priori depend on
$N$).
The Hamiltonian acts on the symmetric space $\H^N=\bigotimes_{\text{sym}}^N
L^2(\R^3)$.
The underlying physical space, if not specified otherwise, is assumed to be
three-dimensional. 

When the external potential $V$ is turned off, $\Psi_{N,0}$ is no longer a
ground state of the Hamiltonian ${H_N\equiv H_N^{V=0}}$ and the time evolution
 $$\Psi_{N,t}=e^{-\i tH_N} \Psi_{N,0}$$
is observed. Although the Schr\"odinger evolution  is linear, its complexity
increases dramatically when $N$ becomes large.
In typical experiments $N$ can be of order $10^5$.
Therefore, for computational purposes, it is important to derive effective
descriptions for collective behaviour of the quantum system.  
In this contribution we review results concerning the approximation of the
time-evolved bosonic many-body quantum system. 

\subsection{Scaling regimes}
From a mathematical point of view, the large number of particles involved will
be modeled by taking the limit $N\to\infty$.
Note that the coupling constant $1/(N-1)$ in front of the interaction is to
ensure the kinetic energy is of the same order (in $N$) as the interaction
energy.
We could choose $1/N$ instead of $1/(N - 1)$ as well.
The interaction potential $w_N$, 
$$w_N (x)=N^{3\beta}w(N^\beta x)$$
for some $w\in L^1(\R^3)$, is chosen to be of $O(1)$ (as $N\to\infty$) in the
sense that $\int w_N =\int w$.
Here and thereafter, if not specified differently, the integration sign will
correspond to integration over the whole space $\R^3$ for each variable.

The parameter $\beta$ in the definition of $w_N$ characterises different
scaling regimes that correspond to different physical situations.
When $\beta=0$ (the so-called Hartree or mean-field scaling) then the
Hamiltonian models a situation in which there are many but weak collisions
between the particles.
This is because the size of the interaction potential is $O(N^{-1})$ (because
of the prefactor $1/(N - 1)$ in front of the interaction term), while the
range of the interaction (which is $O(1)$) is much larger than the mean
inter-particle distance $N^{-1/3}$.
Therefore, heuristically, each particle `sees' all other particles.  

When $\beta>0$, then $w_N$ converges formally to the Dirac-delta interaction
\begin{equation} \label{eq:interactionlimit}
    \biggl(\int w\biggr)\delta_0.
\end{equation}
As long as ${0\le \beta<1/3}$, the range of the interaction potential (which
is $O(N^{-\beta})$) is much larger than the average distance between the
particles and there are many but weak collisions.
Therefore, to the leading order, the interaction potential experienced by each
particle can be still approximated by the effective mean-field potential
$\rho*w_N$ where $\rho$ is the density of the system and $f*g$ denotes the
convolution of two function $f$ and $g$.
If ${\beta > 1/3}$, then the analysis is expected to be more complicated due
to strong correlations between particles.
Despite the physical difference between the case when ${\beta \leq 1/3}$ and
${\beta> 1/3}$, the formal limiting behaviour of the interaction in
\eqref{eq:interactionlimit} is the same in both cases.
Therefore, we will call the regime when ${\beta\in (0,1)}$ the NLS regime as the
limiting effective description of the condensate will be in that case be given
by the nonlinear Schr\"odinger (NLS) equation.  
 
The case when ${\beta=1}$ corresponds to the celebrated Gross--Pitaevskii
regime in which strong correlations occur on very short length scales.
The macroscopic properties of the system are well captured by the famous
Gross--Pitaevskii theory \cite{Gro-61,Pit-61}.
In this theory, a quantum particle is effectively felt by the others as a hard
sphere whose radius is the  scattering length of the interaction potential.
Recall that the scattering length $\ao$ of the potential $w$ is defined by the
variational formula  
\begin{equation}\label{eq:var scat}
  8\pi \ao = \inf\left\{ \int \Bigl(2 |\nabla f| ^2 + w |f| ^2\Bigr) ,
    \quad \lim_{|x|\to \infty}f(x)=1 \right\}. 
\end{equation}
When $w$ is sufficiently smooth, \eqref{eq:var scat} has a minimizer
${0\le f\le 1}$ that satisfies  
\begin{equation}\label{eq:scat-intro-1}
(-2\Delta+w)f=0. 
\end{equation}
The scattering length can then be recovered from the formula 
\begin{equation}\label{eq:scat-intro-2}
 8\pi \ao = \int wf. 
\end{equation}
By scaling, the scattering length of $w_N=N^2w(N\cdot)$ is $\ao N^{-1}$.
If we formally replace the interaction potential $w_N(x-y)$ in $H_N$ by
$8\pi \ao N^{-1}\delta_0(x-y)$, we obtain a Hamiltonian with a Delta
interaction.
Such an object is (in three dimensions) mathematically not well defined.
Nevertheless, it is usually taken as a starting point in the physics
literature on cold gases.

\subsection{Types of approximation}
Recall that our goal is to understand how the $N$-body wave function
$\Psi_{N,t}$ behaves when $N$ is very large.
In the context of dynamics one usually considers three possible effective
descriptions.
The first one, usually called the \textit{leading order approximation},
considers the approximation of $\Psi_{N,t}$ in terms of reduced density
matrices.
Recall, that the (one-body) reduced  density matrix of a state
${\Psi_N\in\gH^N}$ (here we will restrict ourselves to zero temperature) is
the positive, trace class operator ${\gamma_{\Psi_N}:\gH\to \gH}$ with kernel 
\begin{equation} \label{eq:def-1pdm-HN}
  \gamma_{\Psi_N}(x,y) = N \int \d x_{2}\cdots \d x_{N}\,
  \Psi_N (x,x_2,\ldots , x_N)\overline{\Psi_N(y,x_2,\ldots,x_N)}.\quad
\end{equation}
The knowledge of $\gamma_{\Psi_N}$ allows to determine the expectation values
of one-body observables in the state $\Psi_N$.
Indeed, let ${O: \gH \to \gH}$ be an observable and let $O_i$ denote the
corresponding operator acting on the $i$-th particle in the $N$-body space.
Then
\begin{equation} \label{eq:1pdmexpectation}
  \Big\langle \Psi_N, \left(\sum_{i=1}^N O_i\right) \Psi_N\Big\rangle
  =\Tr( O\gamma_{\Psi_N}).
\end{equation}
We will say that the full many-body evolution $\Psi_{N,t}$ is to leading order
approximated by $\Phi_{N,t}$ if 
\begin{equation}  \label{eq:defleadingorder}
\lim_{N\to \infty} \frac1N \Tr | \gamma_{\Psi_{N,t}} -\gamma_{\Phi_{N,t}}|=0.
\end{equation}
Note that, due to \eqref{eq:1pdmexpectation}, the trace norm topology is
natural in this context.
The hope is that $\Phi_{N,t}$ can be determined in an easier way than
$\Psi_{N,t}$.

The convergence \eqref{eq:defleadingorder} is closely related to the
definition of BEC  \cite{PenOns-56}.
We will say that a system of bosons exhibits BEC in the state
${\Psi_N\in\gH^N}$ if  
\begin{equation}  \label{eq:defBEC}
  \lim_{N\to \infty} \Tr \Bigl| \frac1N \gamma_{\Psi_N}
  - |\phi\rangle \langle \phi|\Bigr| =0
\end{equation}
for some ${\phi\in \gH}$.
One then often says that $\phi$ is the wave function of the condensate.
This terminology is related to the fact that if one considers the so-called
Hartree or product state, i.e., the uncorrelated $N$-body wave function of the
form
${\phi^{\otimes N}:=\phi(x_1)\ldots\phi(x_N)}$ in which all particles occupy
the same one-particle state, then
$$\gamma_{  \phi^{\otimes N}}=N |\phi\rangle \langle \phi|.$$
This is why one might sometimes run across the notation
$$``\,\Psi_N \approx \phi^{\otimes N}  \quad \text{to leading order''}$$
which means asymptotic equality in terms of reduced density matrices in the
trace norm topology.
Note, that BEC does not mean that all particles occupy one single-particle
state, but only that a macroscopic fraction does.
In fact, while a product state is a ground state of non-interacting system
(${w_N=0}$), it can't be one of an interacting system.
In particular, if one considers the state (let us skip the symmetry of the
wave function for a moment)  
$$ \Xi_N:= \prod_{i=1}^{N-1}\phi (x_i)\phi^{\perp}(x_N), \quad
\text{with}\quad  \phi^{\perp} \perp \phi $$ 
then obviously 
$$\Xi_N \perp \phi^{\otimes N} \quad \text{in} \,\, L^2(\R^{3N})$$
but \textit{physically}, for large $N$, both states describe a very similar
situation.
In particular, both states exhibit BEC with the same condensate wave function.

Often,  the knowledge about correlations in the system is crucial in order to
understand some of the physical properties of the system.
As seen in the example above, the leading order approximation is not enough
for this purpose.
This is why one considers the most straightforward indicator of closeness:
\textit{the $N$-particle Hilbert space norm}
(or, shortly, \textit{norm approximation}).
More precisely, the goal is to find a $N$-body wave function $\Xi_{N,t} \in
\gH^N$ that is easier to compute than $\Psi_{N,t}$ and such that  
\begin{equation} \label{eq:Nparticlenormapprox}
\lim_{N\to \infty} \|\Psi_{N,t}-\Xi_{N,t}\|_{\gH^N} =0.
\end{equation}
Clearly, since
\begin{equation} \label{eq:normimpliesreduced}
\Tr | O(\gamma_{\Psi_N} -\gamma_{\Xi_N})|\leq 2 \|O\| \|\Psi_N - \Xi_N\|
\end{equation}
for any bounded operator $O$, the norm approximation implies the convergence
of reduced one-body density matrices.  

Another possible way of approximating the wave function is given by the
\textit{Fock space approximation}.
In this approach, one considers the problem in the grand-canonical setting,
where the number of particles in the system is not fixed.
To this end one introduces the Fock space
$$ \cF \equiv \cF(\gH)= \bigoplus_{n=0}^\infty \gH^n
= \C \oplus \gH \oplus \gH^2 \oplus \cdots.$$
The wave function in the Fock space is  denoted by ${\Psi \in \cF(\gH)}$ and 
$$\Psi=\{\Psi^{(0)}, \Psi^{(1)},\ldots, \Psi^{(j)}, \ldots\}$$
where ${\Psi^{(0)}\in \C}$ and ${\Psi^{(i)}\in \gH^i}$ for ${i\geq 1}$.
The inner product on $\cF$ is defined as
$$\langle \Psi_1, \Psi_2\rangle_{\cF}
=\sum_{i\geq 0}\langle \Psi_1^{(i)}, \Psi_2^{(i)}\rangle_{\gH^i}.$$
A state $\Psi_N$ with exactly $N$ particles is described on the Fock space
$\cF$ by a sequence ${\Psi=\{\Psi^{(n)}\}_{n\geq 0}}$ where ${\Psi^{(n)}=0}$
for all ${n\neq N}$ and ${\Psi^{(N)}=\Psi_N}$.  

One can lift the many-body evolution to the Fock space.
To this end we define the Hamiltonian $\cH_N$ on $\cF$ by
\begin{equation} \label{eq:FocksapceHamacion}
(\cH_N \Psi )^{(n)}=\cH_N^{(n)}\psi^{(n)}
\end{equation} 
with the $n$-th sector operator
$$ \cH_N^{(n)}=H_N^V= \sum\limits_{j = 1}^n \bigl(-\Delta_{x_j}+V(x_j)\bigr)
+ \frac{1}{N-1} \sum\limits_{1 \le j < k \le n} {w_N(x_j-x_k)}
$$
where now the subscript $N$ is not related to the number of particles, but
only reflects the scaling in the interaction potential (of course, in the end,
$N$ will also be related with the number of particles in the initial Fock
state; otherwise, there would be no relation with the scaling regime). 

In particular the $N$-particle evolution can be embedded into the Fock space
in the following way 
$$e^{-\i t\cH_N}\{0,0,\ldots, \Psi_N, 0,\ldots\}
  =\{0,0,\ldots, e^{-\i t H_N}\Psi_N, 0,\ldots\}.$$ 
This follows from the fact that the Hamiltonian $\cH_N$ commutes with the
particle number operator $\cN$ given by 
\begin{equation}\label{eq:particlebnumberoperator}
(\cN \Psi)^{(n)}=n \Psi^{(n)}.
\end{equation}

Let $\Psi_0\in \cF$ be a state in the Fock space.
We will say that $\Xi_t \in \cF$ approximates the many-body evolution of
$\Psi_0$ in the Fock space if 
\begin{equation}\label{eq:Fockspaceapprox}
\lim_{N\to \infty} \|e^{-\i t\cH_N}\Psi_0 -\Xi_t \|_{\cF}=0.
\end{equation} 

Sometimes, it is possible to get some information on the $N$-particle space
convergence from  convergence in the Fock space.
This approach, however, usually leads to worse estimates than direct methods
on $N$-particle space and often requires additional assumptions on the initial
states.

\subsection{Outline} The paper will organised as follows.
In the next section we will provide a brief overview about the ground state
properties of (trapped) bosonic systems.  
In Section \ref{sec:leadingorder} we will review the existing results on the
leading order approximation.
In Section \ref{sec:normapprox} we will review the literature on the norm
approximation.
In Section \ref{sec:fockspace} we will mention results on the Fock space
approximation.

\section{Ground state properties}  
Recall that we want to consider initial states that are ground states of
Hamiltonians of the form \eqref{eq:HN-V}.
Therefore it makes sense to briefly review some basics facts concerning this
issue. For more details and references we refer to the excellent review
\cite{Rougerie-20}.

\subsection{Leading order approximation for the ground state}
It is widely expected that ground states of trapped systems exhibit (complete)
BEC.
In fact, when $0\le \beta<1$ we have 
\begin{equation}\label{eq:BEC-energy}
  \lim_{N\to \infty}  \left(\inf_{\|\Psi_N \|_{\gH^N}=1}
    \frac{\langle \Psi_N, H_N^V \Psi_N \rangle}{N}
    - \inf_{\|u\|_{\gH}=1}  \cE^{V}_{\rm{H},N}(u)\right)=0
\end{equation}
where
\begin{equation*} \label{eq:def-EH}
 \cE^{V}_{\rm{H},N}(u):=\frac{1}{N} \langle u^{\otimes N}, H_N^V u^{\otimes
   N}\rangle
 = \int \Bigl(|\nabla u|^2 +V|u|^2 + \frac{1}{2} |u|^2 (w_N*|u^2|)\Bigr).
\end{equation*}
Moreover, if the Hartree energy functional $\cE^{V}_{\rm{H},N}(u)$ has a
unique minimizer $u_{\rm H}$, then  the ground state $\Psi_N^V$ of $H_N^V$
condensates on $u_{\rm H}$ in the sense that  
\begin{equation} \label{eq:BEC-pdm}
  \lim_{N\to \infty} \Tr\biggl| \frac{1}{N}\gamma_{\Psi_N^V}
  -|u_{\rm H}  \rangle \langle u_{\rm H}| \biggr|=0. 
\end{equation}

The rigorous justifications for \eqref{eq:BEC-energy} and \eqref{eq:BEC-pdm}
in various specific cases have been given in
\cite{LieLin-63,FanSpoVer-80,BenLie-83,LieYau-87,PetRagVer-89,RagWer-89,Seiringer-11}.
Later, in a series of works \cite{LewNamRou-14,LewNamRou-14c,LewNamRou-14d},
Lewin, Nam and Rougerie provided proofs in a very general setting.
Most recently, the next order term in the expansion \eqref{eq:BEC-pdm} has
been established in \cite{NamNap-20,BPS-20}. 

When ${\beta=1}$ (the Gross--Pitaevskii regime), the Hartree functional has to
be modified to capture the strong correlation between particles.
In that case 
\begin{equation}\label{eq:BEC-GPenergy}
  \lim_{N\to \infty}  \left(\inf_{\|\Psi_N \|_{\gH^N}=1}
    \frac{\langle \Psi_N, H_N^V \Psi_N \rangle}{N}
    - \inf_{\|u\|_{\gH}=1}  \cE^{V}_{\rm{GP}}(u)\right)=0
\end{equation}
where
\begin{equation} \label{eq:def-EGP}
\cE^{V}_{\rm{GP}}(u): = \int\Bigl( |\nabla u|^2 +V|u|^2 + 4\pi \ao |u|^4\Bigr)
\end{equation}
is the Gross--Pitaevskii functional.
In that case one also has BEC on the Gross--Pitaevskii minimizer
\begin{equation} \label{eq:BECGP-pdm}
  \lim_{N\to \infty} \frac{1}{N}\gamma_{\Psi_N^V}
  =|u_{\rm GP}  \rangle \langle u_{\rm GP}| 
\end{equation}
in trace norm. The convergence \eqref{eq:BEC-GPenergy} has been first proven
by Lieb, Seiringer and Yngvason in \cite{LieSeiYng-00} while
\eqref{eq:BECGP-pdm} has been first proven by Lieb and Seiringer in
\cite{LieSei-02,LieSei-06} (see also \cite{NamRouSei-15}).
More recently, the optimal rates of convergence for \eqref{eq:BEC-GPenergy}
and \eqref{eq:BECGP-pdm} have been given in
\cite{BocBreCenSch-18,BocBreCenSch-20} (translation invariant case) and
\cite{NamNapRicTri-20} (trapped case with smallness condition on $\ao$, see
also \cite{Hainzl-20}).

\subsection{Second order correction} 
The next order correction to the lower eigenvalues and eigenfunctions of
$H_N^V$ is predicted by Bogoliubov's approximation \cite{Bogoliubov-47b}.
In the mean-field limit, this has been first derived rigorously by Seiringer
in \cite{Seiringer-11}, and then extended in various directions in
\cite{GreSei-13,LewNamSerSol-15,DerNap-13,NamSei-15}.
Bogoliubov theory is formulated in the Fock space $\cF$.
At this point, let us briefly recall the notion of second quantization. 

We define the creation operator $a^*(f)$ and the annihilation operator $a(f)$
that for every ${f\in \gH}$ is given by 
\begin{align*}
  \bigl(a^* (f) \Psi \bigr)(x_1,\dots,x_{n+1})
  &= \frac{1}{\sqrt{n+1}} \scalebox{0.90}{$\displaystyle\sum_{j=1}^{n+1}$}
    f(x_j)\Psi(x_1,\dots,x_{j-1},x_{j+1},\dots, x_{n+1}), \\
  \bigl(a(f) \Psi \bigr)(x_1,\dots,x_{n-1})
  &= \sqrt{n} \int \d x_n \, \overline{f(x_n)}\,\Psi(x_1,\dots,x_n)
\end{align*}
for all ${\Psi\in \gH^n}$ and for all $n$.
These operators satisfy the canonical commutation relations (CCR)
\begin{equation} \label{eq:ccr}
  \bigl[a(f),a(g)\bigr]=\bigl[a^*(f),a^*(g)\bigr]=0,
  \quad \bigl[a(f), a^* (g)\bigr]= \langle f, g \rangle
\end{equation}
for all ${f,g\in \gH}$.
Creation and annihilation operators are  used to represent many-body states
and operators on the Fock space.
It is well-known  (see e.g. \cite{Berezin-66} or \cite{Solovej-ESI-2014}) that 
for a symmetric operator $H$ on $\gH$ and an orthonormal basis
${\{f_n\}_{n\ge 1}\subset D(h)}$ of $\gH$ one has
\begin{equation}  \label{eq:second-quantization-H}
  \dGamma (H) := 0\oplus  \bigoplus_{N=1}^\infty \sum_{j=1}^N H_j
  = \sum_{m,n\ge 1} \langle f_m,H f_n\rangle a^*(f_m)a(f_n).
\end{equation}
Similarly, for a symmetric operator $W$  on $\gH \otimes \gH$ such that
$$\langle f_m\otimes f_n, W \,f_p\otimes f_q \rangle
= \langle f_n\otimes f_m, W \,f_p\otimes f_q \rangle$$
for all ${m,n,p,q \ge 1}$ we have 
\begin{align} \label{eq:second-quantization-W}
  &\; 0\oplus 0 \oplus \bigoplus_{N=2}^{\infty} \sum_{1\le i<j \le N} W_{ij}
    \nonumber\\
  =&\; \frac{1}{2} \sum_{m,n,p,q\ge 1} \langle f_m\otimes f_n, W \,f_p
     \otimes f_q \rangle_{\gH^2} \,\,a^*(f_m)a^*(f_n)a(f_p) a(f_q).
\end{align} 

If one does not want to work on a specific orthonormal basis, it is possible
to use the operator-valued distributions $a_x^*$ and $a_x$, with ${x\in\R^3}$,
defined by 
$$
a^*(f)=\int\d x\,  f(x) a_x^*  \qquad \text{and}
\qquad a(f)=\int\d x\, \overline{f(x)}\, a_x
$$
for all ${f\in \gH}$.
The canonical commutation relations \eqref{eq:ccr} then imply that
\begin{equation} \label{eq:ccr-x}
[a^*_x,a^*_y]=[a_x,a_y]=0 
\qquad \text{and}\qquad [a_x,a^*_y]=\delta(x-y).
\end{equation}
The second quantization formulas \eqref{eq:second-quantization-H} and
\eqref{eq:second-quantization-W} can be rewritten as 
\begin{align}  \label{eq:second-quantization-Hxy}
\dGamma (H) &= \int\d x \,\d y\, H(x,y)a^*_x a_y, \\ \hspace*{-10pt}
 \label{eq:second-quantization-Wxy}
  \quad 0\oplus 0 \oplus \bigoplus_{N=2}^{\infty}
  \sum_{1\le i<j \le N}\!\!W_{ij}
            &= \frac{1}{2} \int\d x \,\d y\,\d x'\, \d y'\,
              W(x,y;x',y') a^*_x a^*_y a_{x'} a_{y'},
\end{align} 
where $H(x,y)$ and $W(x,y;x',y')$ are the kernels of $H$ and $W$, respectively.

For example, the aforementioned particle number operator can be written as 
$$
\cN := \dGamma(1) = \bigoplus_{n=0}^\infty n 1_{\gH^n}
= \int\d x\, a_x^* a_x 
$$
and the $N$-body Hamiltonian $H_N$ can be extended to an operator on Fock
space $\cF(\gH)$ as  
\begin{equation}
\cH_N = \dGamma(-\Delta)+\frac{1}{2(N-1)}\int \d x \,\d y\,
w_N(x-y)a^*_x a^*_y a_x a_y . \label{def:2ndquantHamilt}
\end{equation}

As already mentioned, Bogoliubov theory is formulated in the Fock space $\cF$
or, more precisely, the excited Fock space
${\cF(\gH_+)\equiv \cF(\{u_{\rm H}\}^\bot)}$.
Let $\{u_m\}_{m\geq 0}$ be an orthonormal basis of $\gH$ such that
${u_{\rm H}\equiv u_0}$.  
In the mean-field limit, the condensate is described by the Hartree minimizer
$u_{\rm H}$ and the excited particles are effectively described by a quadratic
Hamiltonian $\bH^V$ of the form   
\begin{equation}\label{eq:BogoHamstatic}
\begin{aligned}
 \bH^V&= \sum_{m,n\geq 1}\bigl\langle u_m,(h+K_1)u_n\bigr\rangle a^*_m a_n   \\
 &\quad \mbox{}+ \sum_{m,n\geq 1} \frac12
 \langle u_m\otimes u_n, K_2\rangle a^*_m a^*_n
 +\frac12 \langle K_2,u_m\otimes u_n\rangle a_m a_n
\end{aligned}
\end{equation}
acting on $\cF(\{u_0\}^\bot)$ and  where ${K_1: \gH_+\to \gH_+}$ and
${K_2: \overline{\gH_+} \to \gH_+}$ are operators defined by
\begin{equation}\label{def:K_1,K_2static}
\begin{aligned}
  \langle u, K_1 v\rangle &=\int\d x\,\d y\,
  \overline{u(x)}v(y)u_0(x) \overline{u_0(y)}w(x-y) , \\
  \langle u, K_2 \overline{v}\rangle &=\int\d x\,\d y\,
  \overline{u(x)v(y)}u_0(x) u_0(y)w(x-y) 
\end{aligned}
\end{equation}  
for all ${u,v \in \H_+}$.
Finally, $h$ is the one-body operator given by
$$h=-\Delta+V+|u_0|^2\ast w-\mu$$ 
which comes from the Hartree equation (the Euler--Lagrange equation for the
Hartree functional).
Here $\mu$ is an appropriate constant to make $h u_0=0$.

It has been proven in \cite{LewNamSerSol-15} by Lewin, Nam, Serfaty, Solovej
that if the Hartree minimizer $u_{\rm H}$ is non-degenerate (in the sense that
the Hessian of $\cE_{\rm H}^V(u)$ at $u_{\rm H}$ is bigger than a positive
constant), then the ground state $\Psi_N^V$ of $H_N^V$ admits the norm
approximation 
\begin{equation} 
\label{eq:GS-norm}
  \lim_{N\to \infty} \left\| \Psi_N^V - \sum_{n=0}^N u_{\rm H}^{\otimes (N-n)}
  \otimes_s \psi_n   \right\|_{\gH^N} =0
\end{equation}
where ${\Phi^V = (\psi_n)_{n=0}^\infty \in\cF(\{u_{\rm }\}^\bot)}$ is the
(unique) ground state of $\bH^V$.
Note that the norm convergence \eqref{eq:GS-norm} shows what we mentioned
before, that is, the fact that if ${w\not\equiv 0}$, then $\Phi^V$ is not the
vacuum ${\Omega:=1\oplus 0 \oplus 0 \cdots}$, and hence $\Psi_N^V$ is
\emph{never} close to $u_{\rm H}^{\otimes N}$ in norm.   
For ${\beta>0}$, Bogoliubov theory has been justified for translation invariant
systems in \cite{BocBreCenSch-20a}  (${\beta < 1}$) and in
\cite{BocBreCenSch-19} (${\beta=1}$).

\section{Leading order approximation}\label{sec:leadingorder}
In this section we shall review the results about the leading order
approximation for the Schr\"odinger evolution of a bosonic  many-body wave
function in the sense of \eqref{eq:defleadingorder}.
From a physics perspective we want to answer the following question:
if the initial state of a trapped system exhibits BEC, does the condensate
endure once the trap is switched off and the system starts evolving in time? 

Thus, one would like to show that if the initial many-body wave function
$\Psi_{N,0}$ satisfies 
\begin{equation}  \label{eq:leadinginitial}
  \lim_{N\to \infty} \Tr\, \biggl| \frac1N \gamma_{\Psi_{N,0}}
  - |\phi_0\rangle \langle \phi_0|\biggr| =0
\end{equation}
for some ${\phi\in L^2(\R^3)}$, then 
\begin{equation}  \label{eq:leadingconv}
  \lim_{N\to \infty} \Tr\, \biggl| \frac1N \gamma_{\Psi_{N,t}}
  - |\phi_t\rangle \langle \phi_t |\biggr| =0
\end{equation} 
for some $\phi_t$ which can be found via an effective theory.

First results of that type have been obtained in the mean-field regime
(${\beta=0}$) by Hepp \cite{Hepp-74} (for differentiable  $w$) and by Spohn in
\cite{Spohn-80}  (for  $w$  bounded) (although the setup there, especially in
the work of Hepp, was a priori quite different than the one presented here).  
In the more familiar setup explained in the introduction the discussed
question has been raised in the literature again in the 2000's.
Since then, a substantial effort of the community led to many interesting
results which often differ only slightly.
Those differences might be difficult to spot for non-specialists and one of
the goals of this work is to clarify some of these issues. 

\vspace*{0.4\baselineskip}

\subsection{Results for different scaling regimes} 
\subsubsection{Mean-field scaling}
As mentioned in the introduction, the simplest regime to consider is the
Hartree scaling.
In this case, the general (and imprecise) form of the statement describing the
leading order approximation is given by the following  
\begin{theorem}\label{thm:mfleading}
{\normalfont(Leading order approximation for mean-field dynamics)}
Let $\Psi_{N,0}$ be an initial state satisfying 
\begin{equation}  \label{eq:initialstatemf}
  \lim_{N\to \infty} \Tr \,\biggl| \frac1N \gamma_{\Psi_{N,0}}
  - |u_0\rangle \langle u_0|\biggr| =0
\end{equation}
for a normalized wave function ${u_0\in L^2(\R^3)}$.
Then 
\begin{equation}  \label{eq:ledingorderconvmf}
  \lim_{N\to \infty} \Tr\, \biggl| \frac1N \gamma_{\Psi_{N,t}}
  - |u_t\rangle \langle u_t|\biggr| =0
\end{equation}
where ${\Psi_{N,t}=e^{-\i t H_N}\Psi_{N,0}}$ is the many-body wave function
evolved by the mean-field (with ${\beta=0}$) Hamiltonian $H_N$ (with ${V=0}$)
and $u_t$ is the solution of the time-dependent Hartree equation 
\begin{equation} \label{eq:Hartree-equation}
 \i\partial_t u_t =  \bigl(-\Delta +w*|u_t|^2 -\mu_t\bigr) u_t 
\end{equation}
with  the initial datum $u_0$ and for some appropriate phase ${\mu_t\in \R}$. 
\end{theorem}

\noindent%
Note, that for the leading order the phase plays no role as it does not alter
the projection $|u_t\rangle \langle u_t|$. 
  
The first result of the form of Theorem \ref{thm:mfleading} was obtained by
Bardos, Golse and Mauser in \cite{BarGolMau-00} (with the additional
condition $\langle \Psi_{N,0},H_N \Psi_{N,0}\rangle\leq CN$).
Shortly afterwards  Erd\"os  and Yau obtained in \cite{ErdYau-01} the same
result for  initial states  that had to be a product state.
Clearly, the assumption \eqref{eq:initialstatemf} allows for more general
initial states.
We refer to \cite{BarErdGolMauYau-02} for a recap and comparison of the two
papers.
We note that the work of Erd\"os and Yau allowed to take $w(x)=1/|x|$,
i.e., the Coulomb potential.
Both these works use the BBGKY (Bogoliubov--Born=-Green--Kirkwood--Yvon)
hierarchy method (cf.\ Section \ref{ssec:methods}).
In particular, the BBGKY method does not give any rates of convergence in
\eqref{eq:ledingorderconvmf}.  

The question of the convergence rate has been first answered by Rodnianski and
Schlein in \cite{RodSch-09}.
Using the method of coherent states (cf.\ Section \ref{ssec:methods}) they
showed that the convergence rate in \eqref{eq:ledingorderconvmf} is of the
form  
$$\frac{C}{\sqrt{N}}e^{Ct}$$
for an initial state that is a product state.
Their work included the Coulomb interaction.
This result has been extended in \cite{KnoPic-10} by Knowles and Pickl to
cover more singular potentials (in the sense of the function $w$ rather than
the scaling which was still mean-field) and initial states that are not
necessarily product states but satisfy the more general condition
\eqref{eq:initialstatemf}.
In their work Knowles and Pickl used a method that was developed by Pickl in
\cite{Pickl-11} (which also provides a relatively simple, quantitative proof
of \eqref{eq:ledingorderconvmf} for nice potentials in the mean-field setting,
see also \cite{FroKnoPiz-07,AmmFalPaw-14,Liard-17,Anapolitanos-11}).   

For nicer (bounded and integrable) interaction potentials Erd\"os and Schlein
proved in \cite{ErdSch-09} an optimal convergence rate 
$$\frac{C}{N}e^{Ct}$$
(again, they assumed factorized initial conditions).
This result has been extended in \cite{CheLee-11} by Chen and Lee to cover
more general potentials and then, together with Schlein, further improved to
cover the Coulomb case \cite{CheLeeSch-11} (see also \cite{Kuz-15}). 
 
In \cite{ElgSch-07,MicSch-12} the convergence \eqref{eq:ledingorderconvmf} was
established for particles with a relativistic dispersion relation (the kinetic
energy $-\Delta$  is in that case replaced by $\sqrt{1-\Delta}$) and with
Coulomb type interaction ${w(x) = \pm 1/|x|}$ (this situation is physically
interesting because it describes systems of gravitating bosons, so called
boson stars, and the related phenomenon of stellar collapse).
These systems have been further studied in \cite{Lee-13} (optimal convergence
rate) and \cite{AnaHot-16} (convergence in Sobolev trace norms).
A detailed analysis of the differences in various results regarding the
leading order convergence of mean-field bosonic systems can be found in
\cite{Hott-18}.  

Further developments in the study of the leading order behaviour of  bosonic
systems include the analysis of the mean-field limit coupled to a
semi-classical limit
\cite{AmmNie-08,AmmNie-09,AmmNie-11,FroGraSch-07,FroKnoSch-09,GolMouPau-16,%
  GolPau-17,LafSaf-20},
compound mean-field system systems
\cite{AnaHotHun-18,MicOlg-17,OliMic-19,Lee-19}, systems with magnetic fields
\cite{Luhrmann-12}, systems with three-body interactions \cite{Lee-20},
central limit type theorems for bosonic mean-field dynamics
\cite{BenKirSch-13,BucSafSch-14,KirRadSch-20}. 
Finally, let us mention that for bounded potentials a systematic, perturbative
way to compute higher order terms in the expansion of
\eqref{eq:ledingorderconvmf} has been developed recently in
\cite{BosPetPicSof-19} (see also \cite{PauPul-19}).

\subsubsection{NLS regime}
The analysis of the dynamics becomes more complicated for positive $\beta$.
For ${\beta\in (0,1)}$ the typical result is of the form 
\begin{theorem}\label{thm:NLSleading}
{\normalfont(Leading order approximation in the NLS regime)}
Let $\Psi_{N,0}$ be an initial state satisfying 
\begin{equation}\label{eq:initialstateNLS}
  \lim_{N\to \infty} \Tr \,\biggl| \frac1N \gamma_{\Psi_{N,0}}
  - |u_0\rangle \langle u_0|\biggr| =0
\end{equation}
for a normalized wave function ${u_0\in H^1(\R^3)}$.
Then 
\begin{equation}\label{eq:ledingorderconvNLS}
  \lim_{N\to \infty} \Tr \,\biggl| \frac1N \gamma_{\Psi_{N,t}}
  - |u_t\rangle \langle u_t |\biggr| =0
\end{equation}
where ${\Psi_N (t)=e^{-\i t H_N}\Psi_{N,0}}$ is the many-body wave function
evolved by the many-body Hamiltonian $H_N$ (with ${V=0}$) in the NLS regime
(${\beta\in (0,1)}$) and $u_t$ is the solution of the time-dependent nonlinear
Schr\"odinger equation 
\begin{equation} \label{eq:NLS-equation}
  \i\partial_t u_t =  \bigl(-\Delta +b_0|u_t|^2 -\mu_t\bigr) u_t 
\end{equation}
with the initial data $u_0$.
Here ${b_0=\int w}$ and $\mu_t\in \R$ is an appropriate  phase.  
\end{theorem}

The first, complete proof of Theorem \ref{thm:NLSleading} has been given by
Erd\"os, Schlein and Yau in \cite{ErdSchYau-07}.
It was valid for ${\beta< 1/2}$ and initial states which are a product state.
Later, in \cite{ErdSchYau-10}, Erd\"os, Schlein and Yau extended this result
to all ${\beta\in (0,1)}$ and the general initial states
\eqref{eq:initialstateNLS}.
Both papers used the BBGKY  approach (cf.\ Section \ref{ssec:methods}).
For $\beta < 1/6$, a similar result (for general initial states and without
the assumption on the positivity of the interaction potential) has been
obtained by Pickl in \cite{Pickl-10}.
This work provided also explicit bounds on the convergence rate.

In one dimension, for ${\beta\in (0,1)}$, the problem has been solved by
Adami, Golse and Teta \cite{AdaGolTet-07} (see also
\cite{Rosenzweig-19,AmmBre-12}).
In two dimensions (on a torus) the problem has been studied (for
${\beta<3/4}$) by Kirkpatrick, Schlein and Stafillani in \cite{KirSchSta-11}
and more recently by Jeblick and Pickl in \cite{JebPic-18} (without the
positivity assumption on the interaction).  
Other results about the leading order approximation in the NLS regime include
lower dimensional systems with attractive interactions
\cite{CheHol-16,CheHol-17a}, systems with three-body interactions
\cite{ChePav-11,CheHol-19}, derivations of lower dimensional dynamics from the
three dimensional problem
\cite{CheHol-13,KelTeu-16,CheHol-17,Bossmann-19,Shen-20}.

\subsubsection{The GP regime.} The leading order approximation problem in the
Gross--Pitaevski regime has been first solved by Erd\"os, Schlein and Yau  in
\cite{ErdSchYau-07} where they proved the following theorem: 
\begin{theorem}\label{thm:gpleading}
{\normalfont(Leading order approximation in the Gross--Pitaevskii regime
  \cite[Thm. 3.1]{ErdSchYau-09})} 
Assume ${w\geq 0}$ is a smooth, even potential that decays sufficiently fast
and has  scattering length $\ao$.
Let $\Psi_{N,0}$ be a family of initial wave functions such that
$$\langle  \Psi_{N,0}, H_N \Psi_{N,0}\rangle\leq CN,$$
which exhibit BEC
\begin{equation} \label{eq:initialstateGP}
  \lim_{N\to \infty} \Tr \,\biggl| \frac1N \gamma_{\Psi_{N,0}}
  - |\varphi_0\rangle \langle \varphi_0|\biggr| =0
\end{equation}
for a normalized wave function ${\varphi_0\in H^1(\R^3)}$.
Then 
\begin{equation}  \label{eq:ledingorderconvGP}
  \lim_{N\to \infty} \Tr \,\biggl| \frac1N \gamma_{\Psi_{N,t}}
  - |\varphi_t\rangle \langle \varphi_t |\biggr| =0
\end{equation} 
where ${\Psi_{N,t}=e^{-\i t H_N}\Psi_{N,0}}$ is the many-body wave function
evolved by many-body Gross--Pitaevskii Hamiltonian $H_N$ (with $V=0$ and
$\beta=1$) and $\varphi_t$ is the solution of the time-dependent
Gross--Pitaevskii equation 
\begin{equation} \label{eq:GP-equation}
  \i\partial_t \varphi_t =  \bigl(-\Delta +8\pi \ao|\varphi_t|^2 )\bigr) \varphi_t 
\end{equation}
with the initial condition $\varphi_0$.
\end{theorem}
 
Earlier, the same authors proved in \cite{ErdSchYau-10} the same result under
the additional assumption that 
\begin{equation} \label{eq:smallnessGPESY}
\sup_{r\geq 0}\bigl\{ r^2 w(r)\bigr\}+\int_0^\infty\d r\,r w(r) 
\end{equation}
is small enough.
To remove this smallness condition Erd\"os, Schlein and Yau used an intrinsic
characterization of the correlation structure in terms of the two-particle
scattering wave operator.
Generally, however, both works \cite{ErdSchYau-10,ErdSchYau-09} were based on
the BBGKY hierarchy method and did not provide any quantitative estimates on
the convergence.  

Explicit bounds on the convergence rate in \eqref{eq:ledingorderconvGP} have
been later obtained by Benedikter, de Oliveira and Schlein in
\cite{BenOliSch-15}, by Pickl in \cite{Pickl-15}, and by Brennecke and Schlein in
\cite{BreSch-19}. 

In the periodic setting on a unit torus partial results in the spirit of
Theorem \ref{thm:gpleading} (with a modified many-body Hamiltonian which had a
cut-off to prevent pair interactions whenever at least three particles come
into a region with diameter much smaller than the typical inter-particle
distance) have been obtained by Erd\"os, Schlein and Yau in
\cite{ErdSchYau-06} and then the problem has been solved by Sohinger in
\cite{Sohinger-15}. 

The two-dimensional problem has been solved by Jeblick, Leopold and Pickl in
\cite{JebLeoPic-19}.
At this point let us stress that the Gross--Pitaevskii scaling in two
dimensions is characterized by the scaling ${w_N(x)=e^{2N} w(e^N x)}$ rather
than ${w_N (x)=N^{2}w(N^2 x)}$.
Other results in this regime include the dimensionally reduced dynamics
\cite{Bossmann-20,BosTeu-19}, dynamics in magnetic fields \cite{Olgiati-17},
dynamics of (pseudo-)spinor systems \cite{MicOlg-17b} and central limit type
theorems for dynamics \cite{Rademacher-20}.

\subsection{Methods} \label{ssec:methods}
In this section we shall very briefly explain the two main approaches to prove
the leading order convergence.
For a more pedagogical introduction we refer to the excellent lecture notes of
Benedikter, Porta and Schlein \cite{BenPorSch-16}.

\subsubsection{The BBGKY hierarchy}
The BBGKY approach is based on the idea of investigating the $k$-body reduced
density matrices rather than the wave function itself.
For a given $k$, the $k$-body reduced density matrix of the state $\Psi_N$  is
the generalization of the one-body reduced density matrix and allows to
compute expectation values of $k$-body operators.
It is defined as the operator $\gamma^{(k)}_{\Psi_N}$ on $\gH^k$ whose kernel satisfies 
\begin{equation}\label{def:kbodydm}
\begin{aligned}
&\binom{N}{k}^{-1} \gamma^{(k)}_{\Psi_N}(x_1,\ldots, x_k ;y_1,\ldots , y_k)= \\
&  \int\!\!\d x_{k+1}\ldots \d x_{N}\, \Psi_N (x_1,\ldots ,x_k, x_{k+1},\ldots, x_N)
\overline{\Psi_N(y_1,\ldots ,y_k, x_{k+1},\ldots,x_N)}.
\end{aligned}
\end{equation} 
Note that by setting ${k=1}$ we recover \eqref{eq:def-1pdm-HN}.
In other words,
$$\gamma^{(k)}_{\Psi_N}=\binom{N}{k} \Tr_{k+1}|\Psi_N \rangle \langle \Psi_N |.$$
Using the Schr\"odinger equation, or, more precisely, the von Neumann
equation, one can obtain a hierarchy of equations for
${\widetilde{\gamma}^{(k)}_{\Psi_N}=\binom{N}{k}^{-1}\gamma^{(k)}_{\Psi_N}}$
of the form\renewcommand{\thefootnote}{\fnsymbol{footnote}}%
\footnote[2]{To be precise, the hierarchy \eqref{eq:bbgky} arises for the $H_N$
  with the coupling constant $N^{-1}$ rather than $(N-1)^{-1}$ in front of the
  interaction term.} \enlargethispage{0.2\baselineskip}
\begin{equation}\label{eq:bbgky}
\begin{aligned}
  \i\partial_t \widetilde{\gamma}^{(k)}_{\Psi_{N,t}}
  &=\sum_{j=1}^k
  \Bigl[-\Delta_{x_j},\widetilde{\gamma}^{(k)}_{\Psi_{N,t}}\Bigr]
  +\frac{1}{N}\sum_{i<j}^k \Bigl[w_N(x_i-x_j),
  \widetilde{\gamma}^{(k)}_{\Psi_{N,t}}\Bigr]\\
  &\quad\mbox{}
  +\frac{N-k}{N}\sum_{j=1}^k \Tr_{k+1}\Bigl[w_N(x_j-x_{k+1}),
  \widetilde{\gamma}^{(k+1)}_{\Psi_{N,t}}\Bigr]
\end{aligned}
\end{equation}  
where we use the convention ${\widetilde{\gamma}^{(N+1)}_{\Psi_{N,t}}=0}$.
Consider the mean-field limit, i.e., ${w_N=w}$.
Taking a formal limit ${N\to \infty}$ one obtains
\begin{equation}\label{eq:bbgkylimit}
  \i\partial_t \widetilde{\gamma}^{(k)}_{\infty,t}
  =\sum_{j=1}^k \Bigl[-\Delta_{x_j},\widetilde{\gamma}^{(k)}_{\infty,t}\Bigr]
  +\sum_{j=1}^k \Tr_{k+1}\Bigl[w(x_j-x_{k+1}),
  \widetilde{\gamma}^{(k+1)}_{\infty,t}\Bigr].
\end{equation} 
In the NLS/GP regime the limiting equation obtained from \eqref{eq:bbgky} is
of the form  
\begin{equation}\label{eq:bbgkylimitNLSGP}
  \i\partial_t \widetilde{\gamma}^{(k)}_{\infty,t}
  =\sum_{j=1}^k \Bigl[-\Delta_{x_j},\widetilde{\gamma}^{(k)}_{\infty,t}\Bigr]
  +\sigma \sum_{j=1}^k \Tr_{k+1}\Bigl[\delta(x_j-x_{k+1}),
  \widetilde{\gamma}^{(k+1)}_{\infty,t}\Bigr].
\end{equation} 
with ${\sigma=\int w}$ for ${\beta\in (0,1)}$ and ${\sigma=8\pi \ao}$ for
${\beta=1}$.  

One can check that \eqref{eq:bbgkylimit}/\eqref{eq:bbgkylimitNLSGP} has a
solution given by the Hartree/NLS(GP) equation, i.e.,  
$$\widetilde{\gamma}^{(k)}_{\infty,t}
=\bigl(|u_t\rangle \langle u_t |\bigr)^{\otimes k}$$
where $u_t$ solves the Hartree/NLS(GP) equation.
This leads to the following strategy of proving results like Theorems
\ref{thm:mfleading}-\ref{thm:gpleading} which consists of three main steps: 
\begin{enumerate}
\item Compactness: one needs to prove compactness of the sequence (in $N$) of
  $\{\widetilde{\gamma}^{(k)}_{\Psi_{N,t}}\}_{k=1}^N$ with respect with an
  appropriate (weak) topology. 
\item Convergence: one needs to characterize limit points of the sequence
  $\{\widetilde{\gamma}^{(k)}_{\Psi_{N,t}}\}_{k=1}^N$ as solutions of
  \eqref{eq:bbgkylimit}/\eqref{eq:bbgkylimitNLSGP}. 
\item Uniqueness: one has to prove the uniqueness of the solution of
  \eqref{eq:bbgkylimit}/\eqref{eq:bbgkylimitNLSGP}. 
\end{enumerate}

Proofs of all these steps can be accomplished in various ways depending on
the details of the model (like the regularity and sign of $w$, initial
conditions etc.).
Compactness is usually achieved via a priori estimates.
The larger $\beta$, the more difficult it is to obtain those a priori
estimates.
The a priori estimates will also determine the functional spaces where the
solutions can live in.
In general, the most difficult step is to prove uniqueness.
In the NLS regime (${\beta<1/2}$) Erd\"os, Schlein and Yau \cite{ErdSchYau-07}
proved uniqueness using Feynman diagrams (for example, in their earlier work
with Elgart \cite{ElgErdSchYau-06} they were not able to show uniqueness).
In \cite{KlaMac-08} Klainerman and Machedon provided an alternative approach
based on appropriate (conjectured) space-time bounds on limit points of
$\{\widetilde{\gamma}^{(k)}_{\Psi_{N,t}}\}_{k=1}^N$.
This set up a program in which various research groups tried to establish
these bounds.
That was first successfully done by Kirkpatrick, Schlein and Staffilani on
$\mathbb{T}^2$ in \cite{KirSchSta-11}.
In $\R^3$ the conjecture for ${\beta<1}$ has been established by X. Chen and
Holmer in \cite{CheHol-16c} (see also \cite{CheHol-16b} and the works by Chen
and Pavlovi\'c \cite{ChePav-10,ChePav-13,ChePav-14,ChePav-14b}).
For ${\beta=1}$ a new proof of uniqueness of the hierarchy has been given by
Chen, Hainzl, Pavlovi\'c and Seiringer in \cite{CheHaiPavSei-15}.
Other recent, related works include \cite{AmmLiaRou-20,HerSoh-16,CheHol-20}.

\subsubsection{Quantitative approaches}
As mentioned earlier, the BBGKY hierarchy approach does not, in general,
provide any convergence rate in \eqref{eq:leadingconv}.
In 2009 Rodnianski and Schlein used the coherent states approach to obtain a
quantitative version of Theorem \ref{thm:mfleading} for the first time.
Their method was inspired by the work of Hepp \cite{Hepp-74} and Ginibre and
Velo \cite{GinVel-79}.
The idea is to consider the problem in the Fock space.
The initial state is a coherent state which  is obtained by applying Weyl's
unitary operator ${W(f)=\exp\bigl(a^*(f)-a(f)\bigr)}$ to the vacuum: 
\begin{equation} \label{eq:coherentstate}
W(f) \Omega = e^{-\norm{f}^2/2}\sum_{n\geq0}\frac{1}{\sqrt{n!}}f^{\otimes n}.
\end{equation}
In particular, each $n$-particle component of this state is a product state.
The Fock space evolution is then governed by $\cH_N$ defined in
\eqref{eq:FocksapceHamacion}.
More precisely, the initial state, in order to model a system of $N$ particles
has to be scaled and is given by $W(\sqrt{N}\phi)\Omega$ where $\phi$
corresponds to the initial data condition \eqref{eq:leadinginitial}.  

For states ${\Psi\in\cF}$ in the Fock space we define the one-body reduced
density matrix of $\Psi$ to be the operator (on $\gH$) with the kernel 
$$\Gamma_\Psi (x;y):= \frac{\langle \Psi, a_x^* a_y \Psi \rangle}
{\langle \Psi, \cN \Psi \rangle}.$$
Clearly, this definition reduces (up to a normalization factor) to
\eqref{eq:def-1pdm-HN} for states with exactly $N$ particles.  

Rodnianski and Schlein proved in \cite{RodSch-09} that in the mean-field
limit, the one-body reduced density matrix $\Gamma_{\Psi_{t}}$ of the state  
$$\Psi_t=e^{\i t \cH_N}W(\sqrt{N}u_0)\Omega$$
satisfies
$$\Tr\,\bigl|\Gamma_{\Psi_{t}}-|u_t\rangle\langle u_t|\bigr|
\leq \frac{Ce^{Kt}}{N}$$
for some constants $C$ and $K$.
Here $u_t$ is the solution of the Hartree  equation with initial condition
$u_0$.  
 
Let us stress again, that in this set-up the state $\Gamma_{\Psi_t}$ depends
on $N$ in the grand-canonical sense: $N$ is the \textit{expected} number of
particles in the initial state.
In particular, the result does not, a priori, cover canonical initial
conditions (which would be states coming from wave functions in $\gH^N$).
However, a nice property of coherent states allows to project the result above
to the $N$-particle sector.
To do this, one uses the following representation of a product state in terms
of coherent states: 
$$\frac{\bigl(a^*(u)\bigr)^N}{\sqrt{N!}}\Omega
=d_N \int_0^{2\pi}\frac{\d\theta}{2\pi}\,e^{\i\theta N}
W\bigl(e^{-\i\theta}\sqrt{N}u\bigr)\Omega$$  
where the constant ${d_N=\sqrt{N!}\,N^{-N/2}e^{-N/2}}$ satisfies
${d_N\approx N^{1/4}}$ for large~$N$.  
For the $N$-particle state 
$$\Psi_{N,t}=e^{\it \cH_N}\frac{\bigl(a^*(u_0)\bigr)^N}{\sqrt{N!}}\Omega$$
one then gets 
\begin{equation*} 
\begin{aligned}
  \gamma_{\Psi_{N,t}}
  &=\frac{d_N^2}{N}\int_0^{2\pi}\frac{\d\theta_1}{2\pi}
  \int_0^{2\pi}\frac{\d\theta_2}{2\pi}\,e^{-\i(\theta_1-\theta_2) N} \\
  &\qquad \enskip \times
  \bigl\langle e^{\i t \cH_N} W\bigl(e^{-\i\theta_1}\sqrt{N}u_0\bigr)\Omega,
  a_x^*a_y e^{\i t \cH_N} W\bigl(e^{-\i\theta_2}\sqrt{N}u_0\bigr)\Omega
  \bigr\rangle_{\cF}. 
  \end{aligned}
\end{equation*}
For the inner product in the Fock space one can then use the results about
$\Gamma_{\Psi_t}$.
We see that the price one pays by projecting the Fock space result onto the
$N$-particles sector is given by the constant $d_N^2\approx N^{1/2}$.
This is the reason why the rate of convergence in the canonical ensemble is
$N^{-1/2}$ rather than the $N^{-1}$ above.
In the Gross--Pitaevskii ($\beta=1$) regime a similar approach has been
adopted by Benedikter, De Oliveira and Schlein in \cite{BenOliSch-15} where
they proved (by introducing Bogoliubov transformations to track the
correlations) that
$$\Tr\,\bigl|\Gamma_{\Psi_t}-|\varphi_t\rangle\langle \varphi_t|\big|
\leq \frac{Ce^{Kt}}{N^{1/2}}$$ 
where $\varphi_t$ solves the Gross--Pitaevski equation.
As before, this result was formulated in the Fock space but this time for
\textit{correlated} initial states (cf.\ \eqref{eq:initialFockcorrelated}).
In this case the extension to initial $N$-particle states can only be done
under additional assumptions which also make the convergence slower (see
\cite{ErdMicSch-09} for an earlier analysis on how correlations form). 
Most recently, the method of \cite{BenOliSch-15} has been further extended in
\cite{BreSch-19} by Brennecke and Schlein to  the case of $N$-particle initial
states (with convergence rate $O(N^{-1/2})$).  
 
A different method has been introduced by Pickl in \cite{Pickl-10}.
It has been used in many works, in particular in \cite{Pickl-15} where the
Gross--Pitaevskii regime was analyzed.
The Pickl method, as it now often called, is based on the analysis of a
certain functional, usually called $\alpha_N(\Psi_N, \varphi)$, which counts
(in a weighted way) the number of particles of the $N$-body state $\Psi_N$
that are in the one-particle state $\varphi$.
The functional is applied to the Schr\"odinger time evolved many-body wave
function $\Psi_{N,t}$ and the relevant one-body state $u_t$ in the mean-field
or NLS regime or $\varphi_t$ in the GP regime.
The crucial property of the functional $\alpha_N$ is that
$$ \bigl( \alpha_N(\Psi_N, \varphi)\to 0\bigr)
\Rightarrow\bigl( N^{-1}\gamma_{\Psi_N}\to|\varphi\rangle\langle\varphi|\bigr)$$
as ${N\to\infty}$.
The analysis then concentrates on deriving an estimate on the time derivative
of  $\alpha_N(\Psi_{N,t}, \varphi_t)$ in order to apply Gr\"onwall's
argument.
In particular, this method is suited to cover initial conditions of the
(general) form \eqref{eq:leadinginitial}.

\section{Norm approximation} \label{sec:normapprox}
We shall now review results concerning the norm approximation of many-boson
dynamics.
As mentioned in the Introduction (recall \eqref{eq:normimpliesreduced}), this
notion of closeness is  more precise than the leading order approximation
discussed in the previous section.
The norm approximation is also well suited for initial states that are
$N$-particle states.  

In the mean-field regime, this problem has been first analyzed by Lewin, Nam
and Schlein in \cite{LewNamSch-15}.
They considered the $N$-particle initial states of the form
\begin{equation} \label{eq:PsiN0-intro}
\Psi_{N,0} = \sum_{n=0}^N u_0^{\otimes (N-n)} \otimes_s \psi_{n,0}
\end{equation}
where ${\Phi_0:=(\psi_{n,0})_{n=0}^\infty \in \cF(\{u_0\}^\bot)}$.
This form is motivated by the ground state property \eqref{eq:GS-norm} of
trapped systems.
It was proved in \cite{LewNamSch-15} that when $\beta=0$, the time evolution
$\Psi_{N,t}=e^{-\i tH_N}\Psi_{N,0}$ satisfies the norm approximation 
\begin{equation} \label{eq:PsiN-intro}
  \lim_{N\to \infty} \left\| \Psi_{N,t}
  - \sum_{n=0}^N u_t^{\otimes (N-n)} \otimes_s \psi_{n,t} \right\|_{\gH^N} =0
\end{equation}
where $u_t$ is the Hartree evolution \eqref{eq:Hartree-equation}  with the
right phase factor 
$$
\mu_t=\frac12\int\d x\,\d y\,|u_t(x)|^2 w(x-y)|u_t(y)|^2
$$
and the evolution of ${\Phi_t=(\psi_{n,t})_{n=0}^\infty\in\cF(\{u_t\}^\bot)}$
is generated by a quadratic Bogoliubov Hamiltonian. 
This approach was later developed by Nam and the author of this article in
\cite{NamNap-17} for $\beta<1/3$ and in \cite{NamNap-17a} for
${\beta<1/2}$ (see also \cite{NamNap-17b}). 
 
The crucial ingredient to pass from the $N$-body Hilbert space to the Fock
space (which is natural when describing correlations) is the mapping first
introduced in the static case by Lewin, Nam, Serfaty and Solovej in the
derivation of Bogoliubov theory \cite{LewNamSerSol-15}.
The transformation allows to factor out the condensate from the many-body wave
function and is given by  
\begin{equation} \label{eq:defU}
\begin{array}{cccl}
  U_{N}(t): & \gH^N & \to &
 \displaystyle \cF_+^{\le N}(t):=\bigoplus_{n=0}^N \gH_+(t)^n ,\\[1.2ex]
 & \Psi & \mapsto & \psi_0\oplus \psi_1 \oplus\cdots \oplus \psi_N.
\end{array}
\end{equation}
where ${\gH_+ (t)=\{u_t\}^\perp}$.
The space $\cF_+^{\le N}$ is often called the truncated (because the
$m$-particle sectors for ${m>N}$ are zero), excited (because it describes
excitations around the condensate) Fock space.
The idea is to reformulate the Schr\"odinger evolution
${\Psi_{N,t}=e^{-\i tH_N} \Psi_{N,0}}$ in terms of
$$\Phi_{N,t}:=U_N(t) \Psi_{N,t}$$
which belongs to $\cF_+^{\le N}(t)$ and satisfies the equation
\begin{equation} \label{eq:eq-PhiNt}
\left\{
\begin{aligned}
  \i \partial_t \Phi_{N,t}  &=  \widetilde H_N (t)   \Phi_{N,t}, \\
 \Phi_{N,0} & = 1^{\le N} \Phi_0.
\end{aligned}
\right.
\end{equation}
Here $1^{\le N}$ is the projection onto
${\cF^{\le N}=\mathbb{C} \oplus \gH \oplus \cdots \oplus \gH^N}$ and 
$$
\widetilde H_N (t)= 1^{\le N}
\Biggl[\bH(t)+\frac{1}{2}\sum_{j=0}^4(R_{j} + R_j^*)\Biggr] 1^{\le N}
$$
with 
\begin{align*} 
  &\bH(t):= \dGamma\bigl(h(t)\bigr)
    + \frac12\int\d x\,\d y\,
    \Bigl(K_2(t,x,y)a^*_x a^*_y +\overline{K_2(t,x,y)}a_x a_y\Bigr), \\
  & h(t)=-\Delta+\bigl|u_t(\cdot)\bigr|^2\ast w_N
    -\mu_{t} + Q(t) \widetilde{K}_1(t) Q(t), \\
& K_2(t, \cdot, \cdot)=Q(t)\otimes Q(t)\widetilde{K}_2(t, \cdot, \cdot). 
\end{align*}
Here $\widetilde{K}_1(t)$ is the operator on $\gH$ with kernel
$\widetilde{K}_1(t,x,y)=u_t(x){w_N(x-y)}$  
$\overline{u_t(y)}$, and ${\widetilde{K}_2(t,x,y)=u_t(x)w_N(x-y)u_t(y)}$. 

The state $\Phi_{N,t}$ describes the excitations around the condensate.
Bogoliubov theory assumes that the operators $R_j$ (which we did not write out
explicitly) are small in an appropriate sense.
Thus one may expect that the evolution $\Phi_{N,t}$ in \eqref{eq:eq-PhiNt} is
close (in norm) to the solution of the effective Bogoliubov equation 
\begin{equation} \label{eq:eq-Phit}
\left\{
\begin{aligned}
  \i \partial_t \Phi_t &=  \bH(t) \Phi_t, \\
   \Phi_{t=0} &= \Phi_0.
\end{aligned}
\right.
\end{equation}
Heuristically, the final steps consist of the proof that the number of
(excited) particles in the state $\Phi_t$ is uniformly bounded.
To obtain these kind of bounds one exploits the Bogoliubov equation
\eqref{eq:eq-Phit}.
In particular, to do so, one uses that appropriate norms of the solution of
the Hartree equation (which for ${\beta>0}$ is $N$-dependent) are uniform in
$N$.
This approach, at least for ${\beta<1/3}$, turns out to work also in the case
when the interaction is attractive (up to times for which the effective
equation is well posed).
This has been later exploited by Nam and the author of this note in
\cite{NamNap-19} where they derived the focusing NLS in dimensions one and
two. 

In regimes with ${\beta > 1/2}$ the short scale correlation structure
developed by the solution of the many-body Schr\"odinger equation cannot be
appropriately described by a time-dependent Bogoliubov transformation
satisfying an equation of the form \eqref{eq:eq-Phit}.
A modified approach is needed and this has been done by Brennecke, Nam,
Schlein an the author of this note in \cite{BreNamNapSch-19}.
To take correlations into account more precisely, it is useful to consider the
ground state of the Neumann problem   
\begin{equation}\label{eq:Neum}
  {\left[ -\Delta + \frac{1}{2N} w_N \right]} f_{N}  = \lambda_{N} f_{N}
\end{equation}
on the ball ${|x| \leq \ell}$, for a fixed ${\ell > 0}$.
One fixes ${f_{N} (x) = 1}$, for ${|x| = \ell}$, and extends $f_{N}$ to $\R^3$
requiring that ${f_{N} (x) = 1}$ for all ${|x| \geq \ell}$.
Because of the scaling of the potential $w_N$, the scattering process takes
place in the region ${|x| \ll 1}$; for this reason, the precise choice of
$\ell$ is not very important, as long as $\ell$ is of order one.  

The solution of (\ref{eq:Neum}) can be used, first of all, to give a better
approximation of the evolution of the condensate wave function, replacing the
solution of the limiting nonlinear Schr\"odinger equation
\eqref{eq:NLS-equation} with the solution of the modified, $N$-dependent,
Hartree equation  
\begin{equation}\label{eq:NLSN} 
  \i\partial \varphi_{N,t}
  = -\Delta \varphi_{N,t} + (w_N f_{N} *|\varphi_{N,t}|^2 ) \varphi_{N,t}
\end{equation}
with initial data ${\varphi_{N,0} = \varphi_0}$ describing the condensate at
time ${t=0}$.  

Furthermore, \eqref{eq:Neum} can be used to describe correlations among
particles.
To this end, let
\begin{equation}\label{eq:Bog-trans}
  T_{N,t} = \exp \left( \frac{1}{2} \int \d x\, \d y \,
    \left[  k_{N,t}(x,y) a_x a_y - \text{h.c.}  \right]  \right)
 \end{equation}
with the integral kernel 
\begin{equation}\label{eq:kNt} 
  k_{N,t} (x;y) = (Q_{N,t} \otimes Q_{N,t})
  \Bigl[- N (1-f_N)(x-y) \varphi_{N,t} \bigl((x+y)/2\bigr)^2 \Bigr] 
\end{equation}
where ${Q_{N,t} = 1 - |\varphi_{N,t} \rangle \langle \varphi_{N,t}|}$ is the
orthogonal projection onto the orthogonal complement of the solution of the
modified Hartree equation (\ref{eq:NLSN}).
In particular, in  this context the operator $U_N (t)$ will now project onto
the orthogonal compliment of $\varphi_{N,t}$ and will be denoted by
$U_{\varphi_{N,t}}$.
Since $T_{N,t}$ aims at generating correlations, it is natural to define its
kernel $k_{N,t}$ through the solution of (\ref{eq:Neum}).
In particular, the choice (\ref{eq:Bog-trans}) guarantees a crucial
cancellation in the generator of the fluctuation dynamics.  
The final result can be formulated as follows
\begin{theorem}{\normalfont(Norm approximation in the NLS regime \cite[Theorem
    3]{BreNamNapSch-19})}\label{thm:normbeta<1} 
Consider the initial state ${\Psi_{N,0} \in L^2_s (\R^{3N})}$ with the reduced
one-particle density matrix $\gamma_{\Psi_{N,0}}$ such that 
\begin{equation}\label{eq:ass-gammaN} 
N-\langle \varphi_0, \gamma_{\Psi_{N,0}} \varphi_0 \rangle \leq C   
\end{equation}
and  
\begin{equation}\label{eq:ass-ener}
  \left|\frac{1}{N}\bigl\langle \Psi_{N,0},H_N^V \Psi_{N,0}\bigr\rangle
    -\mathcal{E}_{\rm GP}^V(\varphi_0)\right| \leq C N^{-1}
\end{equation}
with $\mathcal{E}_{\rm GP}^V$ defined in \eqref{eq:def-EGP}.
Let $\Psi_{N,t}$ be the solution of the Schr\"odinger equation with initial
data $\Psi_{N,0}$.
Then, for all ${\alpha < \min\{\beta/2, (1-\beta)/2\}}$, there exists a
constant ${C > 0}$ such that 
\begin{equation} \label{eq:maininfty2}
\begin{split}
  &\;\bigl\|   \Psi_{N,t} -U_{\varphi_{N,t}}^* T_{N,t}^*
  e^{-\i\int_0^t d\tau\; \eta_N (\tau)} \, \mathcal{U}_{2} (t;0)
  \, T_{N,0} \, U_{\varphi_{N,0}} \Psi_{N,0}  \bigr\|^2 \\
  \leq&\;  C N^{-\alpha} \exp(C\exp(C|t|)) 
\end{split}
\end{equation}
for all $N$ sufficiently large and all ${t \in \R}$.
Here $\eta_{N}(t)$ is a phase factor and $\mathcal{U}_{2} (t;0)$ is a unitary
dynamics on $\cF$ with an appropriate quadratic generator that can be defined
using $\varphi_{N,t}, w_N, k_{N,t}$ (see \cite[eq. (41)]{BreNamNapSch-19}).  
\end{theorem}
Notice that the conditions \eqref{eq:ass-gammaN} and \eqref{eq:ass-ener} have
been recently justified in \cite{NamNapRicTri-20}.
Other results on the norm approximation (involving a slightly different
approach  which avoids using second quantization) in the mean-field regime
have been obtained by Mitrouskas, Petrat and Pickl in \cite{MitPetPic-19}.
Petrat, Pickl and Soffer extended this result to a mean-field analysis coupled
to a large volume in \cite{PetPicSof-20}.
A perturbative expansion has been analyzed by Bossmann, Petrat, Pickl and
Soffer in \cite{BosPetPicSof-19}.
In \cite{NamSal-20} Nam and Salzmann provided a norm approximation for systems
with three-body interactions in the NLS regime. 

Let us present an outline of the proof of Theorem \ref{thm:normbeta<1}, as the
strategy is slightly different than the one for ${\beta<1/2}$ and does not
involve the analysis of the Bogoliubov equations. 
The start is similar and involves the action of the map $U_{\ph_{N,t}}$ on
$\Psi_{N,t}$.
This allows us to remove the condensate described at time $t$ by $\ph_{N,t}$
and to focus on the orthogonal fluctuations.
We set 
\begin{equation}\label{eq:phiNt}
\Phi_{N,t} = U_{\varphi_{N,t}}\Psi_{N,t},
\end{equation}
and we observe that ${\Phi_{N,t} \in \cF^{\leq N}_{\perp \ph_{N,t}}}$
satisfies the equation   
\begin{equation} \label{eq:Phi}
\i\partial_t \Phi_{N,t} = \cL_{N,t} \Phi_{N,t}
\end{equation}
with the  generator 
\begin{equation}\label{eq:cLNt}
  \cL_{N,t} = (\i\partial_t U_{\varphi_{N,t}} ) U^*_{N,t}
  + U_{\varphi_{N,t}} H_N U_{\varphi_{N,t}}^*. \end{equation}
A tedious but straightforward computation shows that one can write
\begin{equation} \label{eq:cLNt-2}
 \cL_{N,t}=\sum_{j=0}^4 \cL_{N,t}^{(j)}
\end{equation}
where
\begin{align*}  
  \cL_{N,t}^{(0)}
  &=  \frac{N+1}{2} \product{\varphi_{N,t}}{ \bigl[w_N (1-2f_N)
    \ast|\varphi_{N,t}|^2 \bigr]\varphi_{N,t}} - \mu_{N}(t), \\ 
  \cL_{N,t}^{(1)}
  &=   \frac12\product{\varphi_{N,t}}
    {\bigl[w_N\ast|\varphi_{N,t}|^2\bigr]\varphi_{N,t}}
    \frac{\cN(\cN+1)}{N}  \\ 
  &\quad+\Biggl[ \sqrt{N}\biggl[
    a^*\bigl(Q_{N,t}\bigl[\bigl(w_N (1-f_N)\bigr) \ast|\varphi_{N,t}|^2\bigr]
    \varphi_{N,t}\bigr)   \\
  &  \hphantom{\quad+\Biggl[}\;
    - a^*\bigl(Q_{N,t}\bigl[w_N\ast|\varphi_{N,t}|^2\bigr] \varphi_{N,t}\bigl)
    \frac{\cN}{N}  \biggr] \sqrt{\frac{N-\cN}{N}} +\text{h.c.} \Biggr], \\
  \cL_{N,t}^{(2)}
  &= \dGamma\Bigl(-\Delta+(w_N f_N)*|\varphi_{N,t}|^2+K_{1,N,t}-\mu_{N,t}\Bigr)\\
  &\quad+\dGamma\Bigl(Q_{N,t}\bigl(w_N (1-f_N)
    *|\varphi_{N,t}|^2\bigr)Q_{N,t}\Bigr) \\
  &\quad-\dGamma\Bigl( Q_{N,t}\bigl(w_N *|\varphi_{N,t}|^2\bigr)Q_{N,t}
    + K_{1,N,t}\Bigr)\frac{\cN}{N}\\
  & \quad+ \biggl[\frac{1}{2} \int  \d x\, \d y \,
    K_{2,N,t}(x,y)a^*_x a^*_y  \frac{\sqrt{(N-\cN)(N-\cN-1)}}{N}
    + \text{ h.c.} \biggr], \\
  \cL_{N,t}^{(3)}
  &= \Biggl[\frac{1}{\sqrt{N}}\int \d x\,\d y\,\d x'\,\d y'\,
    (Q_{N,t} \otimes Q_{N,t} w_N Q_{N,t} \otimes 1)(x,y;x',y') \\
  &\hphantom{=\Biggl[\frac{1}{\sqrt{N}}\int}
    \times \varphi_{N,t} (y') a_x^* a_y^* a_{x'}
    \sqrt{\frac{N-\cN}{N}} + \text{h.c.} \Biggr],\\
  \cL_{N,t}^{(4)}
  &= \frac{1}{2N}\int\d x\,\d y\,\d x'\,\d y'\,
    (Q_{N,t}\otimes Q_{N,t}w_N Q_{N,t}\otimes  Q_{N,t}) (x,y;x',y')
     \\&\hphantom{= \frac{1}{2N}\int}\times a^*_x a^*_y a_{x'} a_{y'}
\end{align*}
with
\begin{displaymath} 
  \mu_{N}(t) :=  \product{\varphi_{N,t}}
  { \bigl[\bigl(w_N (1-f_N)\bigr)\ast|\varphi_{N,t}|^2 \bigr]\varphi_{N,t}}
\end{displaymath} 
and
\begin{align*}
K_{1,N,t}&=Q_{N,t}\wt K_{1,N,t}Q_{N,t},\\
K_{2,N,t}&=Q_{N,t}\otimes Q_{N,t} \wt K_{2,N,t}
\end{align*}
where $\wt K_{1,N,t}$ is the operator on $L^2(\R^3)$ with integral kernel
\begin{equation}
  \label{tilde-K1}
\wt K_{1,N,t}(x,y)=\varphi_{N,t}(x) w_N(x-y)\overline{\varphi_{N,t}(y)}  
\end{equation}
and $\wt K_{2,N,t}$ is a function in $L^2(\R^3\times \R^3)$:
\begin{equation}
  \label{tilde-K2}
  \wt K_{2,N,t}(x,y)=\varphi_{N,t}(x)w_N(x-y)\varphi_{N,t}(y).
\end{equation}

\enlargethispage{0.3\baselineskip}

Next, we have to remove the singular correlation structure from $\Phi_{N,t}$.
Since ${\Psi_{N,t} = U^*_{\ph_{N,t}} \Phi_{N,t}}$ and since $U^*_{\ph_{N,t}}$
just adds products of solutions of the nonlinear equation (\ref{eq:NLSN}), it
is clear that all correlations developed by $\Psi_{N,t}$ must be contained in
$\Phi_{N,t}$. 
To remove correlations from $\Phi_{N,t}$ we  apply the Bogoliubov
transformation $T_{N,t}$ defined in (\ref{eq:Bog-trans}).
Unfortunately, $T_{N,t}$ does not preserve the number of particles, and
therefore it does not leave the truncated Fock space $\cF_{\perp
  \ph_{N,t}}^{\leq N}$ invariant.
Since $T_{N,t}$ only creates few particles, this should not be a serious
obstacle.
To circumvent it, it seems natural to give up the restriction on the number of
particles and consider $\Phi_{N,t}$ as a vector in the untruncated Fock space
$\cF_{\perp \ph_{N,t}}$.
The drawback of this approach is the fact that the generator $\cL_{N,t}$
computed in (\ref{eq:cLNt-2}) is defined only on sectors with at most $N$
particles.
So, we proceed as follows; first we approximate $\Phi_{N,t}$ by a new,
modified, fluctuation vector $\wt{\Phi}_{N,t}$, whose dynamics is governed by
a modified generator $\wt{\cL}_{N,t}$ which, on the one hand, is close to
$\cL_{N,t}$ when acting on vectors with a small number of particles and, on
the other hand, is well-defined on the full untruncated Fock space
$\cF_{\perp\ph_{N,t}}$.
To define $\wt{\cL}_{N,t}$ we proceed as follows.
Starting from the expression on the r.h.s.\ of (\ref{eq:cLNt-2}), we replace
first of all the factor $\sqrt{(N-\cN)(N-\cN-1)}$ by $N-\cN$ and then we
replace $\sqrt{N-\cN}$ by $\sqrt{N} G_b(\cN/N)$ where  $G_b(t)$ is the Taylor
series for $\sqrt{1-x}$ around ${x=0}$ up to order $b$.

\enlargethispage{0.1\baselineskip}

Finally, we add a term of the form $C_b e^{C_b |t|}  \cN (\cN/N)^{2b}$ with a
sufficiently large constant $C_b$.
Since the generators $\cL_N$ and $\wt{\cL}_N$ will act on states with small
number of particles, one expects this term to have a negligible effect on the
dynamics (on the other hand, it allows for better control the energy).
With these changes,  one obtains the modified generator
\begin{align} \label{eq:wLNt}
  &\wt\cL_{N,t} = \frac{N+1}{2} \product{\varphi_{N,t}} \notag
  {\bigl[w_N (1-2f_N) \ast|\varphi_{N,t}|^2\bigr]\varphi_{N,t}} - \mu_{N}(t)\\
  &+ \frac12\product{\varphi_{N,t}} \notag
  {\bigl[w_N\ast|\varphi_{N,t}|^2\bigr]\varphi_{N,t} }  \frac{\cN(\cN+1)}{N}  \\
  &+\Bigl[ \sqrt{N} a^*\bigl(Q_{N,t}\bigl[\bigl(w_N (1-f_N)\bigl) \notag
  \ast|\varphi_{N,t}|^2\bigr] \varphi_{N,t}\bigl) G_b(\cN/N)+\text{h.c.}\Bigr]\\
  &-\Bigl[a^*\bigl(Q_{N,t}\bigl[w_N\ast|\varphi_{N,t}|^2\bigr]\varphi_{N,t}\bigr)
  \frac{\cN}{\sqrt{N}} G_b(\cN/N)  + \text{h.c.} \Bigr] \notag \\
  &+\dGamma\Bigl(-\Delta+(w_Nf_N)*|\varphi_{N,t}|^2+K_{1,N,t}-\mu_{N,t}\Bigr)
    \notag\\
  &+\dGamma\Bigl(Q_{N,t}\bigl(w_N (1-f_N) *|\varphi_{N,t}|^2\bigr)Q_{N,t}\Bigr)
    \notag\\
 &-\dGamma\Bigl(Q_{N,t}\bigl(w_N*|\varphi_{N,t}|^2\bigr)Q_{N,t}+K_{1,N,t}\Bigr)
 \frac{\cN}{N} \notag\\
 &+\biggl[\frac{1}{2}\int\d x\,\d y \,K_{2,N,t}(x,y)a^*_xa^*_y
 \frac{N-\cN}{N} + \text{h.c.} \biggr]  \notag\\
 &+\biggl[\frac{1}{\sqrt{N}}\int\d x\,\d y\,\d x'\,\d y'\,
 (Q_{N,t}\otimes Q_{N,t} w_N Q_{N,t}\otimes 1)(x,y;x',y') \notag \\
 & \qquad \qquad \times \varphi_{N,t}(y') a_x^* a_y^* a_{x'}G_b(\cN/N)
 + \text{h.c.}\biggr] \notag \\
 &+\frac{1}{2N}\int\d x\,\d y\,\d x'\,\d y'\,
 (Q_{N,t}\otimes Q_{N,t}w_N Q_{N,t}\otimes  Q_{N,t})(x,y;x',y')
 a^*_x a^*_y a_{x'} a_{y'} \notag \\
 &+ C_b e^{C_b |t|} \, \cN (\cN/N)^{2b}.
\end{align} 
Using this modified generator, we define the modified fluctuation dynamics
$\wt{\Phi}_{N,t}$ as the solution of the Schr\"odinger equation  
\begin{equation} \label{eq:wPhi}
  \i\partial_t \wt\Phi_{N,t} = \wt\cL_{N,t} \wt\Phi_{N,t}, 
\end{equation} 
with the appropriately transformed initial data.
One can then prove that for all ${\alpha < (1-\beta)/2}$, 
there exists a constant ${C > 0}$ such that 
\[ 
  \bigl\| \Phi_{N,t}-\widetilde\Phi_{N,t} \bigr\|^2
  \le C N^{-\alpha} \exp(C\exp(C|t|)) 
\]
for all ${t\in\R}$. 
  
Finally, one applies the Bogoliubov transformation (\ref{eq:Bog-trans}) to the
modified fluctuation evolution $\wt\Phi_{N,t}$ defined in \eqref{eq:wPhi}.
Let 
\begin{equation} \label{eq:defxiNt}
\xi_{N,t}= T_{N,t} \wt\Phi_{N,t}.
\end{equation}
Then ${\xi_{N,t} \in \cF_{\perp \varphi_{N,t}}}$ (with no restriction on the
number of particles) and it solves the Schr\"odinger equation 
\begin{equation} \label{eq:eq-xiN}
  \i\partial_t \xi_{N,t} = \cG_{N,t} \xi_{N,t},  
\end{equation}
with the generator 
\begin{equation} \label{def:wcG}
\cG_{N,t}= (\i\partial_t T_{N,t})T_{N,t}^* + T_{N,t} \wt\cL_{N,t} T_{N,t}^*. 
\end{equation}
As explained above, the application of the Bogoliubov transformation $T_{N,t}$
takes care of correlations and makes it possible for us to approximate the
evolution (\ref{eq:eq-xiN}) with the unitary evolution $\cU_{2,N}$, having as
generator the quadratic part of \eqref{def:wcG}.
The generator $\cU_{2}$  that appears in the statement of Theorem
\ref{thm:normbeta<1} is what one obtains from $\cU_{2,N}$ in limit
${N\to\infty}$.

\section{Fock space approximation} \label{sec:fockspace}
The methods used in the proof of Theorem \ref{thm:normbeta<1} were inspired
strongly by the result of Boccato, Cenatiempo and Schlein who proved in
\cite{BocCenSch-17} an analogous (i.e., for ${\beta<1}$) result in the Fock
space setting (that is in the spirit of \eqref{eq:Fockspaceapprox}).
In that work the authors also used the fluctuation dynamics approach with the
correlations described by the Bogoliubov transformation \eqref{eq:Bog-trans}
(as in \cite{BenOliSch-15}). 

Earlier, the program of deriving effective dynamics in the Fock space setting
for singular interactions was initiated by Grillakis, Machedon and Margetis in
\cite{GriMacMar-10}.
In this work they considered the mean-field regime and proved a result of the
type \eqref{eq:Fockspaceapprox} for Coulomb potentials (see
\cite{GriMacMar-11} for an extension).
In \cite{GriMac-13}, Grillakis and Machedon considered the NLS regime with
${\beta<1/3}$.  

The approach of Grillakis, Machedon and co-authors is in spirit very similar
to the one of \cite{BocCenSch-17}.
However, there is one crucial difference that we would like to point out.
To this end let us briefly explain the approach in \cite{GriMac-13}.

As mentioned before, in the NLS regime correlations play an important role and
to include them in the analysis, similarly to \eqref{eq:Bog-trans}, Grillakis
and Machedon introduce a Bogoliubov transformation (in fact, they did not use
this terminology in \cite{GriMac-13})  
\begin{equation*}
  T(k_t) = \exp \left( \frac{1}{2} \int \d x\,\d y \,
    \left[  \overline{k}_{t}(x,y) a_x a_y - \text{h.c.}  \right]  \right)
\end{equation*}
for \textit{some} function $k_t(x,y)$.
They considered initial Fock space states of the form 
\begin{equation}  \label{eq:initialFockcorrelated}
\Phi(0)=W^*\bigl(\sqrt{N}\varphi_0\bigr)T^*(k_0)\Omega
\end{equation}
which are, a priori, more general than coherent states.
In particular, by choosing $k_0=0$ one obtains a coherent state as an initial
state.
Their idea was to approximate the Fock space many-body evolution 
$$\Phi(t)=e^{-\i t\cH_N}\Phi(0)$$
by an effective quadratic evolution that would capture the creation and
evolution of correlations.
To this end, Grillakis and Machedon postulate that
$$\Phi(t) \approx e^{\i N \xi (t)}W^*\bigl(\sqrt{N}\varphi_t\bigr)
T^*(k_t)\Omega$$
for some phase $\xi(t)$.
Next, they introduce the so-called reduced dynamics 
$$\Phi_{\rm red}(t)=T(k_t)W\bigl(\sqrt{N}\varphi_t\bigr)\Phi(t).$$
Note that ${\Phi_{\rm red}(0)=c\Omega}$ (for some $c$ such that ${|c|=1}$) and
the goal is to find such a $k_t$ so that also the evolved reduced state
satisfies  
\begin{equation} \label{eq:reduceddynamicsGriMac}
\Phi_{\rm red}(t)\approx \Omega.
\end{equation} 
Thus, as we can see, so far the general idea --- the analysis of the
fluctuation dynamics --- is the same as in the work Boccato, Cenatiempo and
Schlein.
Here comes the main difference, however.
While Boccato, Cenatiempo and Schlein postulated the kernel of $k_t$ to be of
the form \eqref{eq:kNt} straight away, Grillakis and Machedon derived an
equation for $k_t$ so that \eqref{eq:reduceddynamicsGriMac} can be satisfied.
More precisely, using the properties of coherent states and Bogoliubov
transformations one can compute (similarly to \eqref{def:wcG}) the time
evolution  
$$\i\partial_t \Phi_{\rm red}(t)=\cH_{\rm red}(t)\Phi_{\rm red}(t)$$
and determine $\cH_{\rm red}(t)$.
The goal is to choose such $\varphi_t$ and $k_t$ so that 
\begin{equation} \label{eq:cHred}
\cH_{\rm red}(t)=N\mu(t)+\int\d x\,\d y\, L_t (x,y) a_x^* a_y + N^{-1/2}\cE(t)
\end{equation}
where $\cE(t)$ is an error term containing polynomials in $a$ and $a^*$ up to
degree four, $L_t (x,y)$ is the kernel of some (self-adjoint) operator and
$\mu(t)$ is an appropriate phase.
This leads to the following set of equations 
\begin{equation} \label{eq:GriMacuncoupled}
\begin{aligned}
  \i\partial_t \varphi_t
  &=  \bigl(-\Delta +w_N *|\varphi_t|^2\bigr)\varphi_t,\\
  \i\partial_t \sh(2k_t)
  &=  -g_N^T\!\circ\! \sh(2k_t)-\sh(2k_t)\!\circ\! g_N
  +m_N\!\circ\! \ch(2k_t)+\ch^T (2k_t)\!\circ\! m_N 
\end{aligned}   
\end{equation}
where $\circ$ denotes the  composition of operators and the operators
$g_N$, $m_N$ are given by  
\begin{equation*}
\begin{aligned}
g_N &=-\Delta+|\varphi_t|^2\ast w_N +\widetilde{K}_1(t),\\
m_N &=\widetilde{K}_2(t).
\end{aligned}
\end{equation*}
Here $\widetilde{K}_1(t)$ and $\widetilde{K}_2(t)$ are the same operators as
in (\ref{tilde-K1}) and (\ref{tilde-K2}), i.e.\ the operators with the kernels
$\widetilde{K}_1(t,x,y)=\varphi_t(x)w_N(x-y)\overline{\varphi}_t (y)$ and
$\widetilde{K}_2(t,x,y)=\varphi_t (x)w_N(x-y)\varphi_t(y)$.
Furthermore, for an operator $A$ with kernel $a$ we define $\sh(A)$ and
$\ch(A)$ to be the operators with the kernels
\begin{align*}
  \sh(a)&:=a+\frac{1}{3!}a\circ\overline{a}\circ a+\cdots,\\
  \ch(a)&:=\delta(x-y)+\frac{1}{2!}\overline{a}\circ a+\cdots,
\end{align*}
respectively.

Thus, as one could expect from  earlier results, the dynamics of $\varphi_t$
is governed by the Hartree equation in the NLS regime.
What is important, the Hartree equation is \textit{uncoupled} from the
equation for $k_t$.
The latter equation is in fact, under appropriate assumptions, equivalent to
the Bogoliubov equations \eqref{eq:eq-Phit} (see \cite{Napiorkowski-18} for
more details).
In \cite{NamNap-17} this equation has been derived in a different manner.
In fact, the derivation of Grillakis and Machedon is closely related to the
diagonalization problem of quadratic Hamiltonians (see \cite{NamNapSol-16} for
more details). 

In the Grillakis--Machedon approach one has to derive the properties of $k_t$
using the equation.
In particular, in order to prove that $\cE(t)$ in \eqref{eq:cHred} can be
treated as an error term, various norms of $k_t$ need to be estimated
uniformly in $N$.
Obtaining such estimates is more difficult in the case of attractive
interactions and this has been done in \cite{Chong-20}.
In \cite{Kuz-17} Kuz extended the analysis to cover the case when
${\beta<1/2}$.
In that work the equations \eqref{eq:GriMacuncoupled} remained unchanged. 

To cover the NLS regime with ${\beta>1/2}$ it turns out that the uncoupled
equations \eqref{eq:GriMacuncoupled} are not sufficient.
In that case, in \cite{GriMac-13b} Grillakis and Machedon suggested a new set
of \textit{coupled} equations that would allow to treat correlations for
larger $\beta$.
Briefly, the equations have been derived from the condition that 
$$ X_1 =0 \quad\text{and}\quad X_2=0$$
where $X_1, X_2$ are one and two-body states in the Fock space given by
$$\cH_{\rm red} \Omega =(X_0, X_1, X_2, X_3, X_4,0,\ldots ).$$
In fact, this condition can be obtained by minimizing the term $X_0$ over
$\varphi_t$ and $k_t$ (this can be seen as a time-dependent version of the
Beliaev theorem introduced in \cite{DerNapSol-13} and used in
\cite{DerMeiNap-13}).
The resulting equations are then quite similar to \eqref{eq:GriMacuncoupled},
but now the operator $m_N$ has to be replaced by the operator $\Theta$ with
the kernel 
$$\Theta (x,y)=-w_N (x-y)\Bigl(\varphi_t (x)\varphi_t (y)
+\frac{1}{2N}\sh(2k_t)(x,y)\Bigr)$$  
and a similar $O\big(\frac{1}{N}\big)$ correction appears in the Hartree
equation.
Under certain smoothness assumptions on $\varphi_0$ and $k_0$, in
\cite{GriMac-17} Grillakis and Machedon were able to show for $\beta\in
(\frac13,\frac23)$ that if $\varphi_t$ and $k_t$ are solution of the coupled
equations, then 
$$\Bigl\|e^{\i t\cH_N}W^*\bigl(\sqrt{N}\varphi_0\bigr)T^*(k_0)\Omega
-e^{\i\chi_t}W^*\bigl(\sqrt{N}\varphi_t\bigr)T^*(k_t)\Omega\Bigr\|_{\cF}
\leq \frac{C}{N^{1/6}}$$
locally in time (i.e., for some ${t\in (0,T_0)}$) and for an appropriate phase
factor $\chi_t$.  

The main difficulty in the proof of the result above lies in the analysis of
the coupled equations, in particular in establishing uniform in $N$ estimates
on the solutions $\varphi_t$ and $k_t$ in certain function spaces.
This result has been further extended for all $\beta<1$ (still locally in
time) in \cite{GriMac-19} and then for all $\beta<1$ but globally in time in
\cite{ChoZha-20} (see also \cite{Chong-18} for results in one dimension).
Finally, let us mention that equations similar to those coupled equations used
by Grillakis and Machedon are often called Hartree--Fock--Bogoliubov equations
and have been analyzed also in \cite{BenSokSol-18,BacBreCheFroSig-16}. 

\bigskip

\noindent\textbf{Acknowledgments.}
The author's work was supported by the National Science Centre (NCN project
Nr. 2016/21/D/ST1/02430).
He would like to thank Phan Th{\`a}nh Nam for useful discussions and remarks
on the initial version of this manuscript and the Institute of Mathematical
Sciences at the National University of Singapore for support, which allowed
him to attend a section of the Workshop on `Density Functionals for Many
Particle Systems' in September 2019.

\end{document}